\documentclass{aa}
\usepackage{graphicx}
\usepackage[varg]{txfonts}
\usepackage{natbib}
\usepackage{multirow}

\def\tex {\ifmmode{{T}_{\rm ex}}\else{$T_{\rm ex}$}\fi}
\def\tmb {\ifmmode{{T}_{\rm mb}}\else{$T_{\rm mb}$}\fi}
\def\ci     {\ifmmode{{\rm C}{\rm \small I}}\else{C\ts {\scriptsize I}}\fi}
\def\hi     {\ifmmode{{\rm H}{\rm \small I}}\else{H\ts {\scriptsize I}}\fi}
\def\hh     {\ifmmode{{\rm H}_2}\else{H$_2$}\fi}

\def\ts     {\thinspace}
\def\kms    {\ifmmode{{\rm \ts km\ts s}^{-1}}\else{\ts km\ts s$^{-1}$}\fi}
\def\msol   {\ifmmode{{\rm M}_{\odot}}\else{M$_{\odot}$}\fi}
\def\lsol   {\ifmmode{{\rm L}_{\odot}}\else{L$_{\odot}$}\fi}
\def\zsol   {\ifmmode{{\rm Z}_{\odot}}\else{Z$_{\odot}$}\fi}

\begin{document}

\title{Probing the merger history of red early-type galaxies
  with their faint stellar substructures}

\author{Brisa Mancillas\inst{1}
\and
P.-A. Duc\inst{2}
\and
F. Combes\inst{1,3}
\and
F.  Bournaud \inst{4}
\and
Eric Emsellem\inst{5,6}
\and
Marie Martig\inst{7}
\and
Leo Michel-Dansac\inst{6}
}
\institute{
Sorbonne Universit\'e, Observatoire de Paris, Universit\'e PSL Univ., CNRS, LERMA, 
F-75014 Paris, France
\and
Observatoire Astronomique de Strasbourg, Universit\'e de Strasbourg, CNRS, 
11 Rue de l'Universit\'e, F-67000 Strasbourg, France
\and
Coll\`ege de France, 11 Pl. Marcelin Berthelot, 75005, Paris, France
\and
Laboratoire AIM Paris-Saclay, CEA/IRFU/SAp, CNRS/INSU, Univ.  Paris Diderot, 91191, Gif-sur-Yvette Cedex, France
\and
European Southern Observatory, Karl-Schwarzschild-Str. 2, 85748, Garching, Germany
\and
Univ. Lyon, Univ. Lyon1, ENS de Lyon, CNRS, Centre de Recherche Astrophysique de Lyon, UMR5574, 69230, Saint-Genis-Laval, France
\and
Astrophysics Research Institute, Liverpool John Moores University, 146 Brownlow Hill, Liverpool L3 5RF, UK
}
   \date{Received  2019/ Accepted  2019}

   \titlerunning{Faint stellar substructures and merger history}
   \authorrunning{Mancillas et al.}

\abstract{Several detailed observations, such as those carried out at the Canada-France-Hawaii Telescope (CFHT),
have revealed prominent Low Surface Brightness (LSB) fine structures that lead to a change in 
the apparent morphology of galaxies. Previous photometry surveys have developed 
observational techniques which make use of the diffuse light
detected in the external regions of galaxies. In these studies, the outer perturbations 
have been identified and classified. These include tidal tails,
stellar streams, and shells. These structures serve as tracers for interacting events and  merging events and  
retain  some memory of the mass assembly of galaxies. Cosmological numerical simulations 
are required to estimate their visibility timescale, among other properties, in order to 
reconstruct the merger history of galaxies. In the present work, we analyze a 
hydrodynamical cosmological simulation 
to build up a comprehensive interpretation of the properties of fine structures. 
We present a census of several types of LSB fine structures compiled using a visual inspection of 
individual snapshots at various points in time. We reconstruct the evolution of the
number of fine structures detected around an early-type galaxy and we compare
it with the merger history of the galaxy. We find that most  fine structures 
are associated with major and intermediate mass merger events. 
Their survival timescale ranges between 0.7 and 4 Gyr. Shells and streams remain
visible for a longer time, while tidal tails have a shorter lifetime. 
These estimates for the survival time of collisional debris provide clues for the interpretation of the
shape and frequency of fine structures observed in deep images with regard to their mass 
assembly. We find that the detectability of stellar streams is most sensitive at the surface 
brightness limit, demonstrating greater visibility at the deepest surface brightness level used in our simulation.
We see between two and three times more streams based on a surface brightness cut of 33 mag arcsec$^{-2}$ than with
29 mag arcsec$^{-2}$.
We find that the detection of shells is strongly dependent upon the projection angle.
}

\maketitle


\section{Introduction}
Within the frame of the Lambda cold dark matter ($\Lambda$CDM) 
cosmological model, galaxy interactions play a major role in hierarchical galaxy 
formation theory.
The models suggest that galaxies are assembled through successive mergers and continuous processes of 
cold gas and dark matter accretion \citep[][]{White1978, Bullock2005, Naab2007, Cooper2013, Cooper2015, Rodriguez-Gomez2016}.
Early studies propose that massive early-type galaxies result from 
a violent relaxation following the merger of two 
equal mass disk galaxies \citep[][]{Toomre1977}. Numerical studies demonstrate that, at high redshifts,
the so-called intermediate-mass mergers of stellar mass-ratio $4:1-10:1$ could be 
progenitors of S0-like galaxies.
In addition, multiple sequential mergers of mass ratios ranging from $4:1-50:1$ or 
even $100:1$ can lead to the build-up of 
elliptical galaxies \citep[][]{Bournaud2005, Bournaud2007,Stewart2008,Karademir2019}. 
The main mode of mass growth for early-type galaxies is also dependent on redshift. 
It is expected that minor 
mergers, along  with stellar accretion, dominate a galaxy's mass growth, rather than major mergers at $z<1$, since they occur 
more frequently \citep[][]{Kaviraj2009,Peirani2010,Hilz2012}. Nonetheless, gas-rich mergers dominate the galaxy mass assembly
in the early formation phases ($2< z< 8$) \citep[e.g.][]{Naab2007}, although direct gas accretion 
from cold streams, as inferred from numerical simulations and, very likely, from observations \citep[e.g.][]{Argudo2016},
may also play a key role in the building of galaxy disks.

Interactions have left vestiges of their existence in the outer regions 
of central galaxies from disrupted satellites and from equal-mass mergers 
\citep[][]{Helmi1999,Ferguson2002,Majewski2003}. This collisional debris is mainly made up of
stellar streams, plumes, tidal tails and
stellar shells \citep[][]{Mihos2005,McConnachie2009, Janowiecki2010, Martinez-Delgado2010, vanDokkum2005, vanDokkum2014}.
The detection of each class of collisional debris is essential since they 
trace the last merger events and store information about the mass 
assembly of their progenitors. In addition, their identification may modify their 
apparent morphology and then their classification
 in optical imaging surveys, such as the Sloan Digital Sky Survey \citep[SDSS][]{York2000}. 

The present work emphasizes a qualitative analysis of Low Surface Brightness (LSB) fine structures identified in a 
numerical simulation conducted in a cosmological context. 
The first imaging observations of collisional debris to exhibit the distinctive shapes 
of tidal tails and stellar streams, mainly in early-type galaxies, were published in the {\it{Atlas of Peculiar Galaxies}} catalog 
\citep[][]{Arp1966}. In the case of shell structures, \cite{Malin1983} reported a catalog of 137 elliptical galaxies exhibiting shells 
or ripple-like structures. On the theoretical side, the first computational models of interacting galaxies revealed that the
perturbations and debris are the outcome of gravity tides in disk galaxies \citep{Toomre1972}.
Since that time, substantial efforts undertaken to develop observational techniques 
aimed at identifying these substructures.
The techniques include, for example, the unsharp masking and amplification of photographic images \citep[][]{Malin1977,Malin1978}, 
sky subtraction of individual images \citep[][]{Mihos2005}, and extraction of structural components with GALFIT \citep[][]{Peng2002}.
Additionally, the application of galactic archeology also allows us to establish a demography of LSB structures in nearby 
galaxies \citep[][]{McConnachie2009, Crnojevic2013}, although its application is restricted to galaxies at a very low redshift due 
to spatial resolution and sensitivity. 

The search for diffuse light was also a task carried out by amateur astronomers 
using   simple cameras 
\citep[e.g.][]{Martinez-Delgado2009}. Their long-exposure observations revealed spectacular images that later inspired the implementation of
new techniques in conventional telescopes \citep[][]{Martinez-Delgado2010}. 
Nonetheless, in spite of these 
innovative techniques, the detection of merger remnants continues to be problematic. Low spatial resolution or a limited 
field of view make it more difficult to recognize tenuous morphologies 
in the outer regions of galaxies. 
Perhaps the main issue is related to low surface brightness, expected to be below the level of
28 mag.arcsec$^2$ and, most frequently, below $\sim$30 mag.arcsec$^2$ 
\citep[e.g.][]{Johnston2008, Cooper2010}. Thanks to the availability and advancement of 
deep imaging programs for nearby 
galaxies, it has been possible to detect the prominent structures that surround galaxies,
with the large field of view provided by the mosaic camera \citep[CFHT/MegaCam, NGVS,][]{Ferrarese2012} or the specialized Dragonfly 
Telephoto Array, designed for LSB astronomy \citep[][]{vanDokkum2014}.

In their pioneering work, \cite[][]{Malin1983} revealed that numerous shells
could be observed as tracers of interactions  and mergers in  137
early-type galaxies \citep[][]{Malin1983}. 
In the survey of 36 field galaxies aimed at detecting  fine structures, 
\citet[][]{Schweizer1985} 
identified 44$\%$ of weak shells and $\sim$ 10$\%$ of plumes and tidal tails. In more recent studies, \citet[][]{Tal2009} 
present an optical image sample of 55 luminous elliptical galaxies and, by using a tidal interaction parameter, they  find that 
73 $\%$ of these displayed tidal features. Meanwhile, \citet[][]{Atkinson2013} report $\sim$ 26 $\%$ of tidal features in their
sample of 1781 galaxies from the CFHT Legacy Survey. On the other hand, \citet[][]{Krajnovic2011} identify 8 $\%$ of tidal 
debris in a sub-sample of the ATLAS$^{3D}$ sample of 260 early-type galaxies, whereas in the massive early-type galaxies of MATLAS deep 
imaging survey, it was found that $\sim$ 16 $\%$ displayed streams and shell-like features, and $\sim$ 22$\%$ showed 
tails and plumes \citep[][]{Duc2017}. In the most recent studies, \citet[][]{Hood2018} report an incidence of 17$\%$ of tidal features
in the 1,048 galaxies of the RESOLVE survey, as well as \citet[][]{Kado-Fong2018}, who identify 18 $\%$ shell galaxies 
and 82 $\%$ stellar stream systems from the subsample of 1,201 galaxies of the SDSS spectroscopy images taken from 
the Hyper Suprime-Cam Subaru Strategic Program (HSC-SSP).

When it comes to numerical studies, only a few works have taken a census of tidal debris thus far. In the analysis of the hydrodynamical 
Illustris simulation, \citet[][]{Pop2018} report that 18 $\%$ of their massive galaxies display shell-like structures as a result 
of merger events with stellar mass ratios $\gtrsim$ 10:1. \citet{Karademir2019} explore a wide parameter space in
mass ratios and relative orbits. They find that streams are formed by satellite infalling with a large angular momentum,
while shells are the result of the radial infall of satellites with low angular momentum. More and more often, fine structures are being
identified and classified through automatic techniques, such as those developed in \cite[][]{Pawlik2016, Hendel2019, Walmsley2019}.
It is still necessary to identify fine structures visually in order to teach the machine how to improve the precision of
image recognition and to verify  procedures. In addition, different dark matter halo models should be
explored since the fine structures will greatly depend on dark matter spatial distribution.

Several scenarios have been proposed to explain the morphology characteristics of tidal features. 
The disruption of merging galaxies may yield different features according
to their relative orbits and the  geometry of the encounter \citep[][]{Amorisco2015, Hendel2015}. 
Analytical and numerical studies show 
that stellar shell structures are the remnants of disrupted satellites 
on near-radial orbits, while stellar streams are generated 
by near-circular orbits \citep[][]{Quinn1984, Dupraz1986, Johnston2008, Karademir2019}.
Collisional debris is made up of transient structures that can last for short times or 
be disappear as a result of posterior interactions in a few 
Gyr \citep[][]{Stewart2008}. Their recognition is primarily dependent 
on their surface brightness, in addition to
several other factors, such as the type of substructure and orientation in the sky \citep[][]{Duc2017}. Thus, the determination 
of their survival timescale is crucial in reconstructing the  merging history 
of galaxies, together with their numerical simulations. 
Until now, numerical works have been used to constrain these timescales.
The earliest studies of semi-analytical models focus in the signatures of satellites 
orbiting the Milky Way. \citet{Johnston1999} reported a "preceding passage" 
of $\sim 0.7$ Gyr for tidal debris for the Sagittarius 
Dwarf Galaxy, demonstrating that the satellite is rapidly being disrupted and 
will only survive a few pericentric passages, that is, $\sim 1.3$ Gyr.
Hybrid models composed of semi-analytical and N-body simulations of Milky Way-type 
stellar halos from \citet{Bullock2005} 
estimate a median accretion time of $\sim 5$ Gyr for their satellite systems.

Hydrodynamical simulations of equal-mass mergers performed 
in isolated environments have determined several
timescales based on different methods. \citet{Lotz2008} report an 
average timescale of $\sim 1.5$ Gyr using the Gini coefficient. Meanwhile,
\citet[][]{Ji2014} use visual inspection to estimate a merger-feature time (the moment
when faint features disappear) of $\sim 1.38$ Gyr 
(for $\mu=25$ $mag.arcsec^2$), which is comparable to the timescale 
computed by \cite{Lotz2008}. Current deep imaging surveys go much deeper and further studies
are needed to estimate updated timescales.

The goal of the present work is to identify and classify the LSB features observed in a host halo based on the hydrodynamical 
simulation of \citet[][]{Martig2009}.  These are zoom-in resimulations  that reveal the fate of a typical
massive galaxies in great detail after having undergone cold gas accretion and several mergers during a Hubble time. The cosmological
context is taken from a previous large-scale simulation with reduced spatial resolution. The high resolution of the zoom-in simulation
is capable of identifying most fine structures occurring at any epoch.
In this analysis, five observers have visually classified a full mock catalog of 
stellar surface brightness maps  (cf. the snapshots in the Appendix). 
The aims of the classification are to:  (i) characterize the shape of the faint 
stellar structures on the outskirts of the 
main central galaxy, (ii) take a census of the number of these structures and 
classify them into tidal tails, streams, 
and shells, (iii) to estimate the survival time of each substructure and study 
their dependence on the surface brightness 
limit and the projection on the sky plane. 
In section 2, we describe the numerical simulation carried out 
in a cosmological context (section 2.1), the high-resolution re-simulation
(section 2.2), and the phases of the galaxy evolution (section 2.3). We describe the catalog of stellar surface brightness maps  in
Section 3. Results are reported in Section 4 and discussed in Section 5.
Section 6 presents a summary of our conclusions.

\section{Numerical simulations}

\subsection{Simulation technique}

To analyze the galaxy evolution in the $\Lambda$CDM cosmological model, 
we use the numerical simulation of \cite{Martig2009}. 
The approach demonstrates how the growth of a massive stellar spheroid 
through a series of mergers can be enough to stabilize a disk which is, by then,
no longer self-gravitating, and therefore,
quench star formation.  Through this process, the galaxy becomes
a red early-type object while continuing
to accrete gas. The technique that was used consists of two steps: first a
$\Lambda$-CDM cosmological simulation is run, with only dark matter,
while the merger and accretion history for a given halo is extracted.
In the second step, the mass assembly history is re-simulated at a
higher resolution, replacing each halo with a realistic galaxy,
containing gas, stars, and dark matter.\\

\subsubsection{Cosmological simulation}

The cosmological simulation used to carry out this study was performed with the RAMSES code
(\cite{Teyssier2002}), based on the Adaptive Mesh Refinement (AMR) technique. 
The box has a comoving length
of 20 $h^{-1}$ Mpc, and contains $512^{3}$ dark matter particles with a mass 
resolution of $6.9 \times 10^{6}$ $M_{\odot}$.
The dark matter halos are identified using the HOP algorithm of \cite{Eisenstein1998}, and the merger and  accretion 
histories are extracted following the main halo from $z=2$ to $z=0$. All the halos (of the merger history) and diffuse particles 
(of the accretion history) that cross a fixed spherical boundary drawn around the target halo are recorded. The sphere is larger 
than the initial main halo at $z=2$ and encompass satellites, as well as "diffuse" particles,
that is, those that don't belong to any halo.
The second step zoom-in resimulations were carried out for Milky Way-mass galaxies in low
density environments \citep{Martig2012}. Within this sample, we selected a case
where there is a major merger with a post-merger early-type phase galaxy.  There is no
particular selection bias, nor were the galaxies selected on the basis of an abundance of
fine substructures.

\subsubsection{High-resolution resimulation}
The next step consisted of a high-resolution resimulation using a particle-mesh 
code described in 
\cite{Bournaud2002} and \cite{Bournaud2003}, in which the gas dynamics is 
modeled using a sticky particle scheme.
The maximal spatial resolution is 130 pc and the mass resolution is $1.4 \times 10^5$ $M_{\odot}$ for stellar particles
initially present in the galaxies, $2.1 \times 10^4$ $M_{\odot}$ for gas and stellar particles formed during the simulation,
and $4.4 \times 10^5$ $M_{\odot}$ for dark matter particles. Star formation is computed following a Schmidt-Kennicutt
Law (\cite{Kennicutt1998}), where the SFR is proportional to the gas density to the exponent of 1.5, setting a threshold
for star formation at $0.003$ $M_{\odot}$ $pc^{-3}$. The simulation does not include supernova explosions, nor
AGN feedback.

\begin{figure*}
\centering
\includegraphics[ scale=2.3, trim=5 0 0 10,clip]{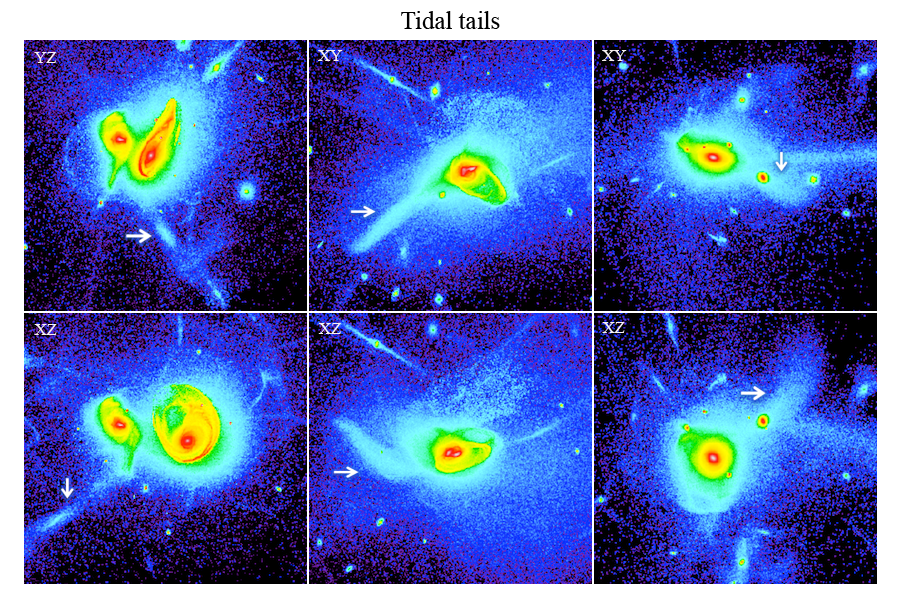}
\caption{Examples of interactions between central galaxy and its nearby companions, 
        which demonstrate protuberant tidal 
tails (indicated with a white arrow) for three different objects 
        (resp., right, middle and left panels) in 
two random projections: XY in the upper panel and XZ at the bottom.}
\label{tidal}
\end{figure*}

The resimulation starts at redshift $z=2$ and evolves down to $z=0$ , including all
the halos and the "diffuse" particles embedded
in the spherical boundary around the target halo, recording their mass, position, velocity and spin. Each halo is replaced
by a realistic galaxy made of gas, stars and dark matter particles, and by replacing each diffuse particle with a blob of lower-mass, 
higher-resolution gas and dark matter particles (see Appendix A of \cite{Martig2009} for more details). Each small halo starts to interact 
with the main halo following the orbital and spin parameters given at the beginning of the cosmological simulation (e.g., Fig \ref{tidal} and subsequent figures). 


\subsection{Merger history and stellar mass growth of the  studied galaxy}
The central galaxy studied in this simulation has a mass similar to that of the Milky Way. Its corresponding host halo does not
belong to a rich group or a galaxy cluster as most early-type galaxies (ETGs) in the MATLAS survey (see snapshots in the Appendix for the galaxy 
environment). During the simulation, the host halo
increases from a mass of $2 \times 10^{11}$ $M_{\odot}$ at $z=2$ to $1.4 \times 10^{12}$ $M_{\odot}$ at $z=0$.
The galaxy evolution can be decomposed into 3 main phases
(see the plot 3 of \cite{Martig2009}). 
The simulation starts out with a gas-rich disk galaxy and an intense phase of minor mergers with a mass ratio 
between $1:4$ and $1:10$. This period lasts from $z=2$ to $z \sim1$. Afterwards, a quiet phase takes place between
$z \simeq 1$ and $z \simeq 0.2$. This is a quiescent period studied by \cite{Martig2009} in which
star formation is suppressed. Over this period, there is no major event occurring, however, diffuse gas  is continuously being accreted.
In the final phase, the galaxy undergoes a major merger event, along with an increase in the diffuse gas accretion
rate, which lasts from  $z=0.2$ to $z=0$.
These three phases are representative of any merging history in the life of a galaxy and, thus,  our estimations of life-time of fine structures will be statistically significant.\\

\begin{figure*}
\centering
\begin{tabular}{c}
\includegraphics[ scale=0.5, trim=6 0 0 0,clip]{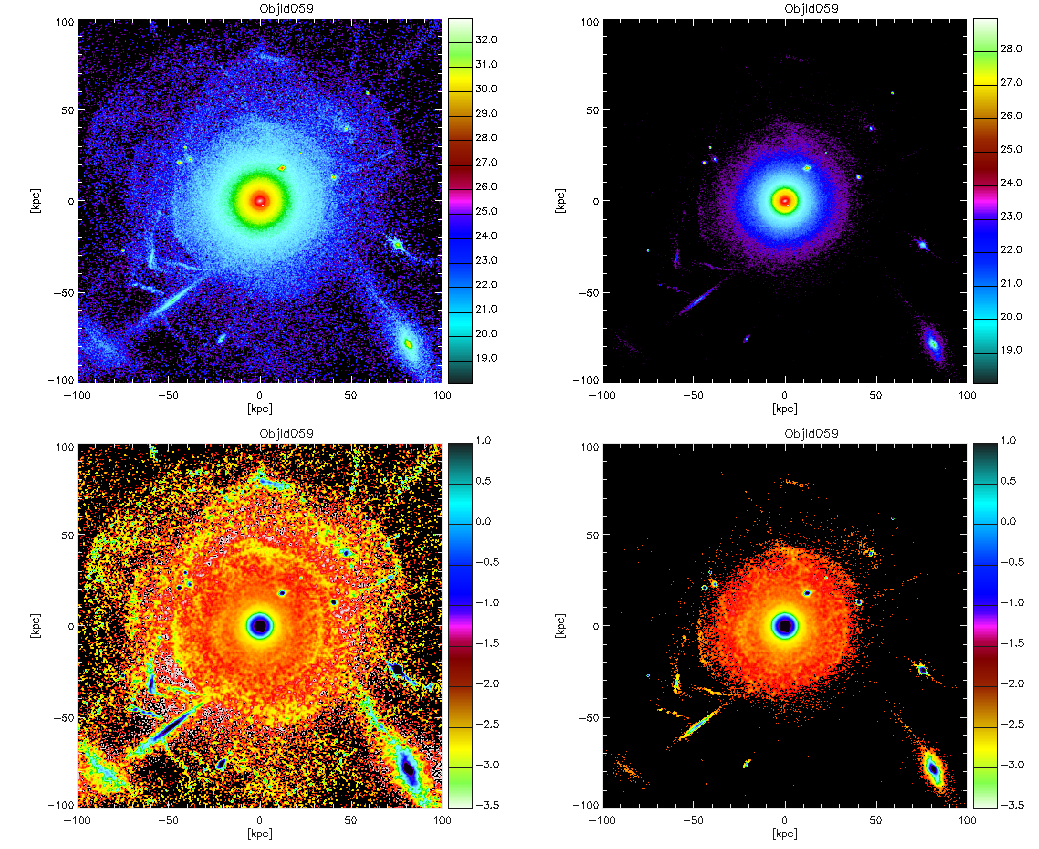}
\end{tabular}
\caption{Surface brightness map of central galaxy shown in g-band with two
        different color scales (resp., upper and lower panels) and with cut at 33 mag arcsec$^{-2}$ (left panels) and 
        29 mag arcsec$^{-2}$ (right panels).}
\label{obj}
\end{figure*}

%
\begin{figure*}
\centering
\begin{tabular}{c}
\includegraphics[ scale=2.3, trim=5 0 0 10,clip]{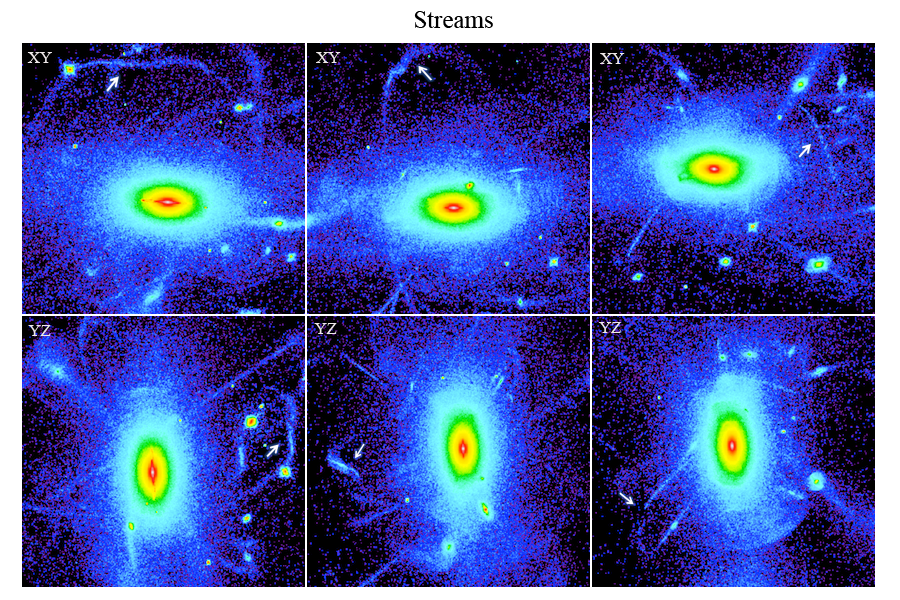}
\end{tabular}
\caption{Examples of central galaxy hosting disrupted satellites based on presence of prominent stellar
 streams (indicated with a white arrow) for three different objects 
        (resp., right, middle and left panels)
 in two random projections: XY in the upper panel, and YZ at the bottom.}
\label{stream}
\end{figure*}
%
\begin{figure*}
\centering
\begin{tabular}{c}
\includegraphics[ scale=2.3, trim=5 0 0 10,clip]{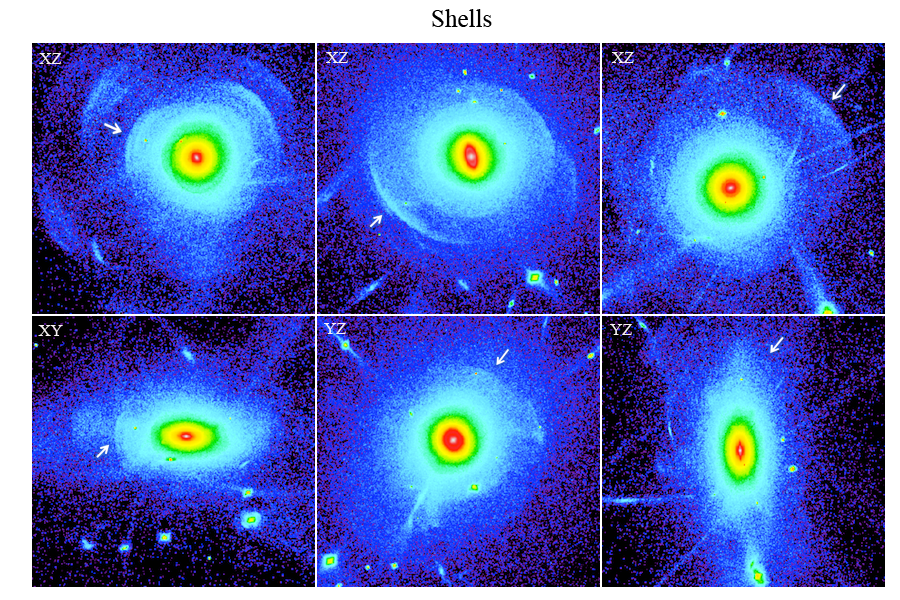}
\end{tabular}
\caption{Examples of  central galaxy displaying several shells around it (indicated by white arrow) for three different objects (resp., right, middle and left panels) 
        in three random projections: XZ in the upper panel, XY and YZ at the bottom.} 
\label{shell}
\end{figure*}
%
%
%
\begin{figure}
\centering
\begin{tabular}{c}
\includegraphics[ scale=0.69, trim=3 0 0 0,clip]{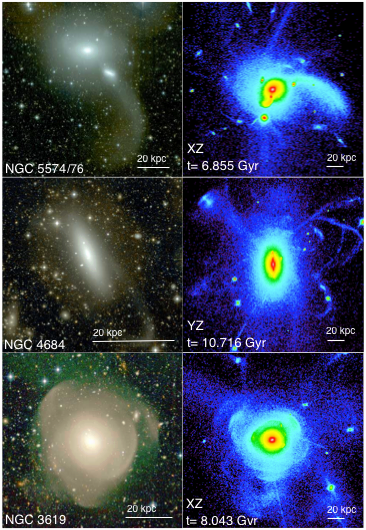}
\end{tabular}
\caption{Comparison of LSB substructures around observed and simulated galaxies.
 \textit{Left panels}:
examples of ETGs from MATLAS Survey. \textit{From top to bottom}: system 
 NGC 5574/76 (tidal tails emanating from
NGC 5576), stellar stream in NGC 4684 and interleaved shells in NGC 3619. 
 \textit{Right panels}: Examples of mock
surface brightness images exhibiting prominent tidal tails, streams, and shells at different times of simulation
in random projections.
        Note the white scale bar indicating 20~kpc: the scales are comparable
        across observations and simulations.}
\label{comparisonONS}
\end{figure}


\section{Surface brightness maps}
\subsection{Mock image generation}
We use the models of \citet{Bruzual2003} (BC03)  to compute mock images of the stars in the simulation. We follow the 
technique developed in \citet{Michel-Dansac2004}. We briefly describe how the mock images are made. For all 
stellar particles in the simulation, we compute the stellar spectra with a linear interpolation of age in the BC03 tables. 
For this, we  use  tables based on a Salpeter IMF \citep{Salpeter1955}. 
We assume a solar metallicity for the stellar 
particles. The simulation does not follow the chemical evolution. For one projection, particles are projected on a 2D grid using 
a Cloud in Cell (CIC) algorithm \citep[][]{Birdsall1969}. Then, for each pixel we sum up the spectra of individual particles to 
get the composite spectrum. 
These spectra are finally integrated in the CFHT/MegaCam photometric filters in order to produce surface brightness maps.

We have compiled a catalog of images for the 35 snapshots of the simulation and the three projections, presenting a total of 105 images. For each image, we prepared two g-band maps: one with a cut at 33 mag arcsec$^{-2}$
\citep[][]{Valls-Gabaud+2017}
and the other at 29 mag arcsec$^{-2}$. The latter value corresponds to the surface brightness limit of the MATLAS survey.
We also added unsharp masked maps, a technique that has proven very 
efficient in enhancing LSB substructures.
The first snapshot starts at 3.5 Gyr and the last one ends at 13.7 Gyr (in addition to an extra simulation 
which evolved until 15.7 Gyr). The time interval is 0.3 Gyr. 
Below, we provide a phenomenological description of collisional debris, based on the observational work 
of \citet{Duc2015}(see e.g., Figure \ref{obj}).
To visualize tidal streams, tails and shells,  we believe that  dust extinction
is not significant  because the fine-structure features are located in
the outer parts of galaxies, where there is little gas or dust.

\subsection{Visual classification}

For the visual identification, we proceeded in a similar way to earlier observational surveys aimed at identifying substructures
optically \citep[e.g.][]{Tal2009, Atkinson2013, Duc2015, Duc2017}.
A team of five members carried out a visual inspection of each mock image. We 
identified and classified the  fine structures that were observed and took a census of them for both surface brightness cuts.
Note that all the images (various snapshots and projections) in the catalog have been randomly ordered for the 
classification in order to avoid two close snapshots in the same projection following the other in the catalog or two projections of the same snapshot. This ensures that our classification is not influenced  by
the stellar structures of recent history or another viewing angle. 
Indeed, in real observations, there is no equivalent of two consecutive snapshots of the 
same merger event. In our case, each image is initially considered entirely 
independently of any other. This also allows us to verify the consistency of the method 
once the merger history  is reconstructed. 

Figures \ref{tidal}, \ref{stream}, and \ref{shell} demonstrate three examples of tidal tails, stellar streams, and shells,
respectively, each projected in two directions, and based on the case of the lowest surface brightness limit, that is, 
$\mu =$33 mag arcsec$^{-2}$ .
  Below, we define the substructures categorized within the three morphologies mentioned above.



Tidal tails are thick, radially elongated structures that are connected to the host galaxy. The visual impression is that the tidal tail 
appears to be  emanating from the galaxy. In the case of any confusion between tail and stream, we relied on the aspect of thinness to 
distinguish between streams and tail; tails being thicker than streams.
 The thickness separation between the two is of the order of 300pc.
Over time, the dilution of tidal tails implies a widening, so there is no confusion.

Stellar streams are thin, elongated stellar  structures which look like narrow, long filaments. They can be associated 
with  galaxy satellites, exhibiting the known S-shaped structures that emanate from the main galaxy and characterize satellite
disruption. But this is not a requisite, so they might also not be associated clearly with any galaxy, or physically associated 
with the central galaxy. 

Shells have circular concentric shapes and are sharp-edge arc structures.
Depending on their intrinsic nature and their projection, they appear to be aligned 
with a common axis or randomly spread around the central galaxy. When aligned to the major
axis of the host, they are interleaved, meaning that they appear to accumulate  
alternatively on  each side of the galaxy. This stellar  accumulation can be
associated with successive apocenters of an oscillating radial orbit following a satellite  accretion event.
When they extend to larger radii,  they become more diffused.

To illustrate the three kinds of fine structures, we compare
typical examples for each, in both observations and simulations,
in Figure \ref{comparisonONS}.


\begin{table}
  \begin{center}
    \caption{Stellar mass ratio of main mergers.}
    \label{tab1}
    \begin{tabular}{|r|c|l|l}\hline
\multicolumn{3}{ |c| }{Properties of main mergers} \\
 \hline
     t$_{merger}$ [Gyr] & M$_{\star}$ [M$_{\sun}$]& stellar mass ratio\\
      \hline
       4.77 & $4.4\ 10^{9}$  & 0.14 (7:1) \\
      5.67 & $13.0\ 10^{9}$ & 0.31 (3:1) \\
      6.26 & $1.5\ 10^{9}$  & 0.04 (25:1) \\
      7.20 & $12.0\ 10^{9}$ & 0.17 (6:1) \\
      7.75 & $5.0\ 10^{9}$  & 0.06 (16:1) \\     
      11.9 & $74.0\ 10^{9}$ & 0.67 (1.5:1) \\
      \hline
    \end{tabular} 
  \end{center}
  \medskip
  Note: stellar mass ratio has been measured just before galaxy merger
  \vspace{1mm}
\end{table}
%
\section{Results}
%
\subsection{Evolution of  stellar mass}

\begin{figure}
        \includegraphics[angle=0,width=7.7cm]{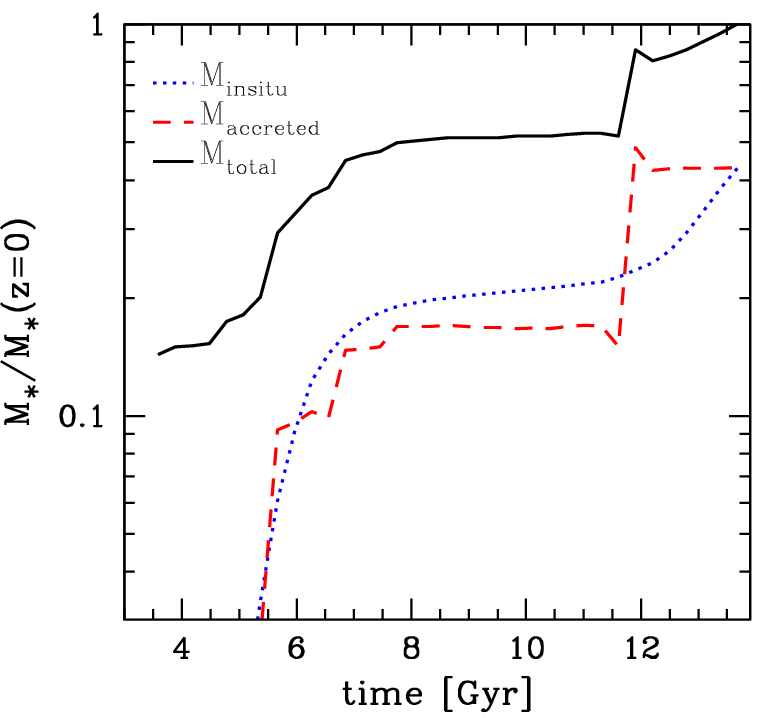}
    \caption{Evolution of  stellar mass for  main galaxy as function of time. The total stellar mass (black solid line) has
     been decomposed into the in situ mass (blue dotted line), that is, stars formed in the main galaxy, and accreted mass (red dashed line),
      that is, stars formed in a companion galaxy and then accreted into the main galaxy.}
    \label{Evol_Ste_Acc_Mass}
\end{figure}

The stellar mass of the central galaxy is plotted as a function of time  in Fig.~\ref{Evol_Ste_Acc_Mass}. We distinguish
the provenance of the stellar mass into two categories. First, we show the contribution of the {\it{in situ mass}}, 
referring to the stars born inside of the main galaxy. Secondly, the {\it{ex situ mass,}} or accreted mass referring to
the stars formed in the satellite galaxies and accreted afterwards into the main galaxy. We note that since the simulation starts 
at $z=2$ with preprepared galaxy models, a third contribution to stellar mass is the initial mass of the galaxy. 
These three phases describe the evolution of the main galaxy. In addition, at larger scales, a number of satellites orbit 
the halo all over the simulation. 
In Table \ref{tab1}, we describe the properties of the most important satellites that interact with the main galaxy during the 
simulation. The merging satellites are identified using an arbitrary number. Their merging time, their stellar
mass, and their mass ratio with respect to the main galaxy are all listed. 

\subsection{Identified substructures}
Having defined the method and described the mock catalog of surface brightness images, we explore  our findings
and the substructure properties observed in this analysis.

\begin{figure}
\centering
\begin{tabular}{c}
\includegraphics[angle=0,width=7.7cm]{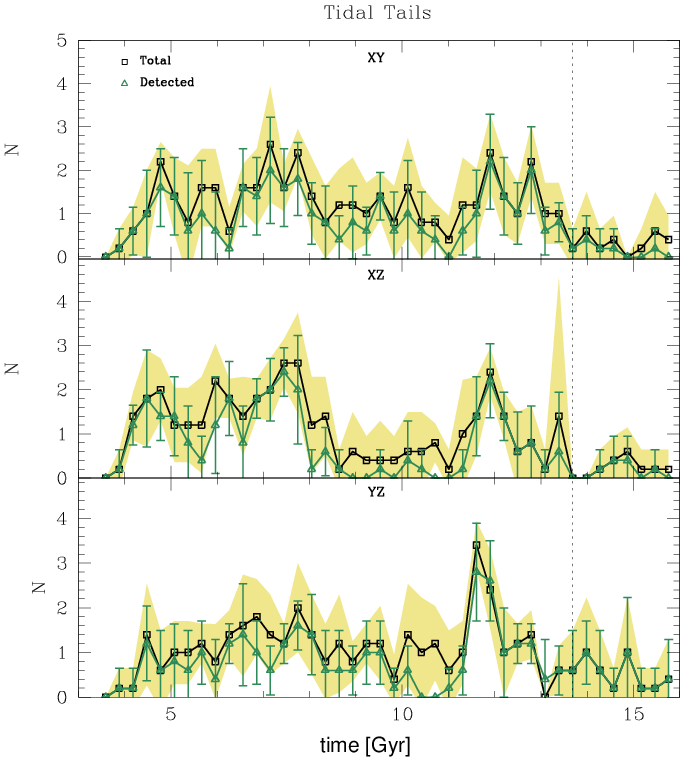}
\end{tabular}
\caption{Evolution of  tidal tail numbers.  The three panels show
        the results for the three projections (XY, XZ, and YZ). In each panel, 
        the black lines and the yellow shaded area represent
the total number of tidal tails and their respective scatter, whereas the green ones are the
    number of tidal tails that are detectable on images with a cut in
    surface brightness at 29 mag arcsec$^{-2}$.
        The time (t=0) corresponds to the Big-Bang, and the re-simulation is longer
        than a Hubble time, beyond z=0, to estimate the future of fine structures.
        }
\label{evotidal}
\end{figure}

\begin{figure}
\centering
\begin{tabular}{c}
\includegraphics[angle=0,width=7.7cm]{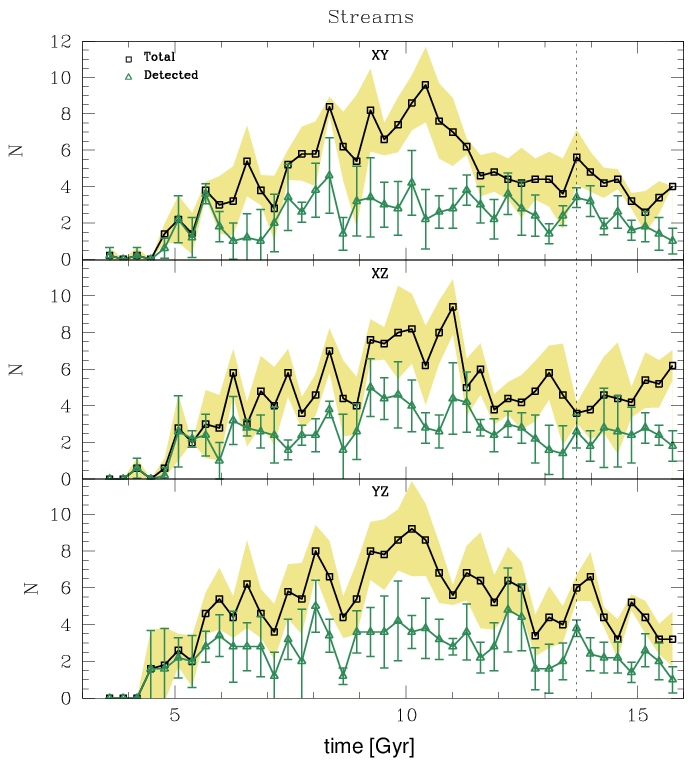}
\end{tabular}
\caption{Evolution of stellar stream numbers.  Three panels show
    results for  three projections. In each panel, the black lines and 
        the yellow shaded area are
    the total number of stellar streams and their respective scatter, 
        whereas the green ones represent the
    number of stellar streams that are detectable on images with a cut in
    surface brightness at 29 mag arcsec$^{-2}$.
        The time (t=0) corresponds to the Big-Bang.}
\label{evostreams}
\end{figure}

In Figures \ref{evotidal}, \ref{evostreams}, and \ref{evoshells}, we present the time evolution of the number of tidal 
tails, stellar streams and shells, respectively, for three different projections, $x-y$, $x-z$ and $y-z$ (as indicated 
in each panel). 
Time begins at the start of the resimulation (z=2), and continues after the present
time (z=0), so that the future of fine structures can be estimated.
The panels display the curves of the average value from the five team members who made the visual 
inspection for each time interval (snapshot) and their respective scatter. We present this census for both 
surface brightness limits $\mu$ = 29 mag arcsec$^{-2}$ (``detected") and $\mu$ = 33 mag arcsec$^{-2}$ (``total"),
respectively.

In addition, in order to analyze  the correlation between the generation of features and the presence of mergers, in Figure 
\ref{ComparisionMergers} we show the accreted stellar mass (red line in upper panel) and we use arrows to indicate where 
the mergers take place. We point out the most significant mergers and their mass ratios with red numbers. In the lower
panel, we present the corresponding evolution of each fine structure.
Here we plot the highest numbers of features over the
three projections for tidal (blue dotted line), streams (green dotted and dashed line) and shells (black dashed 
line). The identification of each substructure detected during the formation of the galaxy corresponds to the
time period of $t=3.5$ Gyr ($z=2$) to $t=16$ Gyr ($z>0$). 
 The major merger at T=11.5 Gyr is destroying most of the previous fine-structures;
  shells in particular. Some more shells and tails are rebuilt after that at the same rate
  as before. It will take some time to reach the same level of fine structures
  as was reached in the quiet phase.

We note that the curves indicating the number of fine structures as a function of time
(Fig \ref{evotidal}, \ref{evostreams}, and \ref{evoshells}) are reconstructed and
reordered afterwards. It is remarkable that they make sense, that is, the correlations between individual consecutive snapshots as visible and as expected. 
Below, we describe the main characteristics
of the census of tidal debris. \\

%
\subsection{Formation mechanism and survival time}

\begin{figure}
\centering
\begin{tabular}{c}
\includegraphics[angle=0,width=7.7cm]{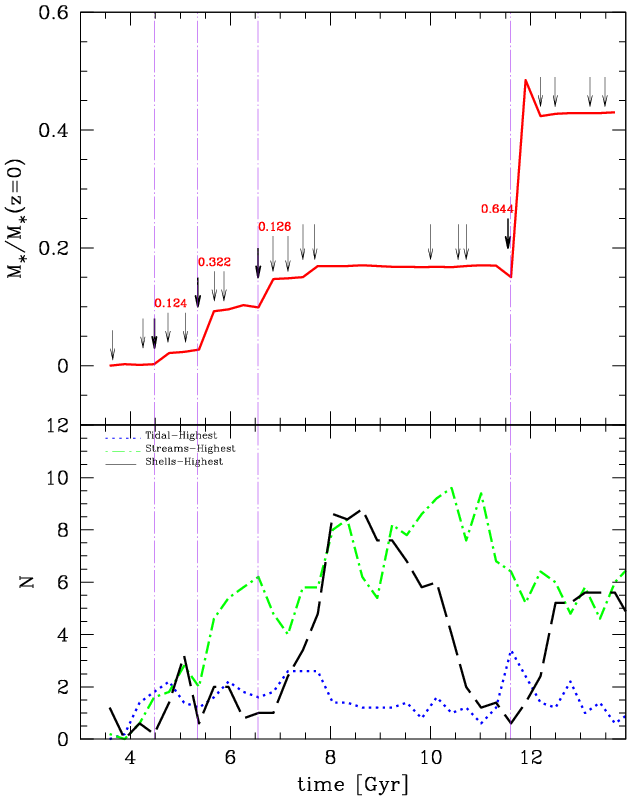}
\end{tabular}
\caption{Comparison between accreted stellar mass and all types of fine structures.
 {\it Upper panel}: red line represents  evolution of accreted mass. 
        Arrows indicate when mergers occur. 
 Red numbers over bold arrows correspond to intermediate mass and major mergers. 
        {\it Lower panel}:
 evolution of highest value of three projections in each fine structure. Blue dotted line is for tidal
 tails, green dotted and dashed line is for streams and black dashed line is for shells. 
        Purple dotted and
 dashed lines in both panels indicate the correlation between main mergers and substructures. }
\label{ComparisionMergers}
\end{figure}

The formation scenario for each substructure is the first bit of information  we can infer from the evolution of the simulation. 
We know by construction all the merger events of the central galaxy in all its evolution phases. 
The counting of the substructures is, thus, a direct outcome and it allows us
to draw correlations between individual substructures and the corresponding merger event. In figure \ref{ComparisionMergers},
we represent the correlation between the generation of features and the occurrence of mergers.\\

\subsubsection{Tidal tails.} We find that tails are produced during the first and the third phase of  galaxy evolution as a
result of tidal forces acting within the host galaxy. In the first phase, we observe an average of two tails associated with the Gyr
intermediate-mass merger and major merger events (mass ratios between 7:1 and 3:1). In the second phase,
their detection drastically decreases due to the lack of any kind of mergers.
In the last phase, a peak of incidence appears when the major 
merger (mass ratio of 1.5:1) takes place. Prominent tidal tails can be evaluated 
through visual inspection at 6.8 Gyr and
11.9 Gyr in the snapshot evolution (see figures \ref{snapXY}, \ref{snapXZ} and \ref{snapYZ} in Appendix A).
We estimate a survival time of $\sim$ 2 Gyr.\\

\subsubsection{Stellar streams.} Streams are the most frequent fine structure features in our simulation 
(an average of $\sim$ 8 streams detected through visual inspection,
see Fig \ref{evostreams}). They appear in the first series of consecutive minor mergers, increasing in number
over time, although some of them are destroyed when the following merger events occur. The highest peak of 
incidence is reached during the quiescent phase (second phase). The absence of important
mergers during this phase allows for the 
increase in the production of streams and their survival time by the lack of processes to that would eliminate them. 
They trace the initial orbit of the satellites which  are disrupted during this quiet phase, producing long, extended
narrow streams. We observe a decrease in the detection number when the major merger
appears and eliminates many of them in the last phase of the simulation.
Under all the physical conditions explained above, we derive a life time of $\sim$ 3 Gyr.

The stellar streams are present during the 
quiescent phase of the galaxy, and even demonstrate their peak in this period, 
while there is no minor merger (nor major one) in all this phase. This 
may look paradoxical. Either the stellar streams develop from the minor
mergers coalesced in the previous phase, or there is another mechanism
to form these streams. The latter could be due 
to the accretion of cold gas \citep[e.g.][]{Bournaud2005}.\\

\subsubsection{Shells.} In our simulation, shell formation is associated with both intermediate-mass mergers and 
major mergers. Previous works have shown they are commonly related with intermediate-mass mergers 
\citep[e.g.][]{Pop2018, Karademir2019}. The largest generation of shells is visible during the quiet phase, just after 
the first phase of mergers has occurred, whereas a second peak of shells is produced just after the major merger takes place.
We attribute this high incidence of shells to the same phenomenon as in the case of streams, that is: when there 
are no violent processes to eliminate them, they remain numerous in the quiet phase.
Following the snapshot evolution (see figure \ref{snapXY}, in the Appendix), we note that once
the corresponding satellite falls into the central potential, shells appear and acquire their known arc-like shape. Initially,
they are spread around the galaxy at small distances from the center and are distributed in an interleaved way. They  
accumulate near the apocenters of their orbits, and develop progressively  to the outskirts of the central galaxy.
Consequently, due to their density wave nature, they propagate towards larger and larger 
distances, and their numbers increase. According to the duration of their incidence peak, that is, from the width of the
curves in Figure \ref{evoshells}, and assuming that no shell is produced during the quiet phase, we estimate 
a lifetime of $\sim$ 4 Gyr.\\

\subsection{Sensitivity on the surface brightness cut}
The surface brightness limit is one of the main issues relevant to the identification of fine features, according to observational studies.
Traditional images from the SDSS survey \citep{York2000} reach a value of 26.4 mag arcsec$^{-2}$ in the g-band.
In the sample from CFHT Legacy Survey \citep{Atkinson2013} and the observations with the MegaCam of \citet{Tal2009}, also in the g-band, they barely detect substructures at $\sim$ 28 mag arcsec$^{-2}$. Deeper observations like
the LSB-optimized NGVS Survey \citep{Ferrarese2012}, the ETG from ATLAS$^{3D}$ \citep{Duc2015}, and the Large Program 
MATLAS Survey \citep{Duc2017} achieve a value of 29 mag arcsec$^{-2}$. These values relate to
the local surface brightness. However, it is possible to define integrated-light surface brightness,
and a very low level of scattered light (one order of magnitude below
the best-performing telescopes), with the lowest 
limit achieved with the Dragonfly Telephoto Array \citep{vanDokkum2014}, having a value of of 32 mag arcsec$^{-2}$. The latter reveals
 spectacular hidden substructures in the external regions of galaxies.
These surface brightness limits could be reached locally by dedicated space-based missions,
such as the Messier project \citep{Valls-Gabaud+2017}.

On the simulation side, there is the freedom to reproduce deep images with low surface brightness.
The values range from 28 mag arcsec$^{-2}$ \citep{Ji2014} to even 38 mag arcsec$^{-2}$ \citep{Bullock2005,Johnston2008}.
Motivated by all these previous studies on the detectability of faint features in merger remnants, the present study
selects two values 
of the surface brightness limits, $\mu$ = 29 mag arcsec$^{-2}$ and $\mu$ = 33 mag arcsec$^{-2}$, respectively, to  take a census
of LSB substructures. 

\subsubsection{Tidal tails.}
We do not find any substantial dependency on the surface brightness for this type of substructure.
We detect a comparable number
of tidal tails for both surface brightness cuts per time step (snapshot).
The curves in the three panels of Figure \ref{evotidal} follow the same 
trend over the course of the evolution. We attribute this result to their prominent nature, which is a result of  major 
interactions and, therefore, fall above the detection threshold.

\subsubsection{Stellar streams.} 
\begin{figure*}
\centering
\begin{tabular}{c}
\includegraphics[ scale=0.515, trim=30 0 0 0,clip]{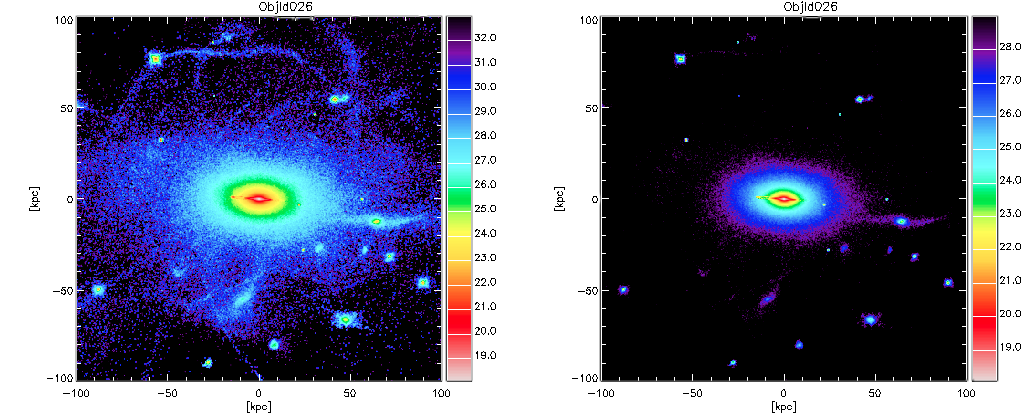}
\end{tabular}
\caption{Example of stellar stream appearance with a cut at 33 mag arcsec$^{-2}$ 
        (left panel) and
        29 mag arcsec$^{-2}$ (right panel). }
\label{SBExampleStream}
\end{figure*}

We find a high dependence on the surface brightness limit for this type of substructure.
As seen in the three panels of figure \ref{evostreams}, the incidence of streams is considerably higher
for $\mu$ = 33 mag arcsec$^{-2}$
than for $\mu$ = 29 mag arcsec$^{-2}$. The peak of incidence reaches an average number of 9 streams for the first cut, 
while for the second cut, the peak only reaches 4 streams. Streams are tracers of minor mergers and, hence, they have a narrow 
morphology, making their detection difficult. They are faint, but as they are more narrow, they may be identifiable for longer than 
the more diffuse tails. Figure \ref{SBExampleStream} presents the surface brightness map of a particular
stream, under the two cuts,  to illustrate a clear example of this difference. With the low level of surface brightness (left panel),
 a wide variety of extended streams surround the central galaxy, along with  the tail of a prominent satellite, in the higher surface brightness cut (right panel), and only the central regions of the whole system are seen. Stellar streams are the type of substructure which present the strongest
 dependency on surface brightness in comparison with tails and shells.

\subsubsection{Shells.} Shells are slightly sensitive to surface brightness. A significant difference is 
observable between the maximal peaks of the two curves in the middle panel of Figure \ref{evoshells} (projection $y-z$). 
For this peak, we detect, on average, two shells more for the case of $\mu$ = 33 mag arcsec$^{-2}$ than for $\mu$ = 29 mag arcsec$^{-2}$.
 
\subsection{Dependence on the projection}
Another important issue relevant to the identification of substructures using visual inspection 
is the effect of the projection. Fortunately, our approach reproduces three projections for each snapshot in order to consider the effects of 
the sky orientation on each class of substructures.

\subsubsection{Tidal tails.} 
with the phrase in bold as this format amounts to the equivalent of bullet points.
We do not find any dependence on the projection. We compare the three 
projections in the corresponding panels of  
Figure \ref{evotidal} and we find a similar tendency for the curves across all periods of time. 
The same number of tidal tails can
be detected in any projection of the simulation.
We attribute this behavior to the 3D nature of tails when encounters are not coplanar.

\subsubsection{Stellar streams.} Streams do not exhibit dependence on the orientation. They are visible 
in all directions. The same number of streams
can be detected in any projection. Curves in the three panels of Figure \ref{evostreams} 
follow the same trend.

\subsubsection{Shells.} 
\begin{figure}
\centering
\begin{tabular}{c}
\includegraphics[angle=0,width=7.7cm]{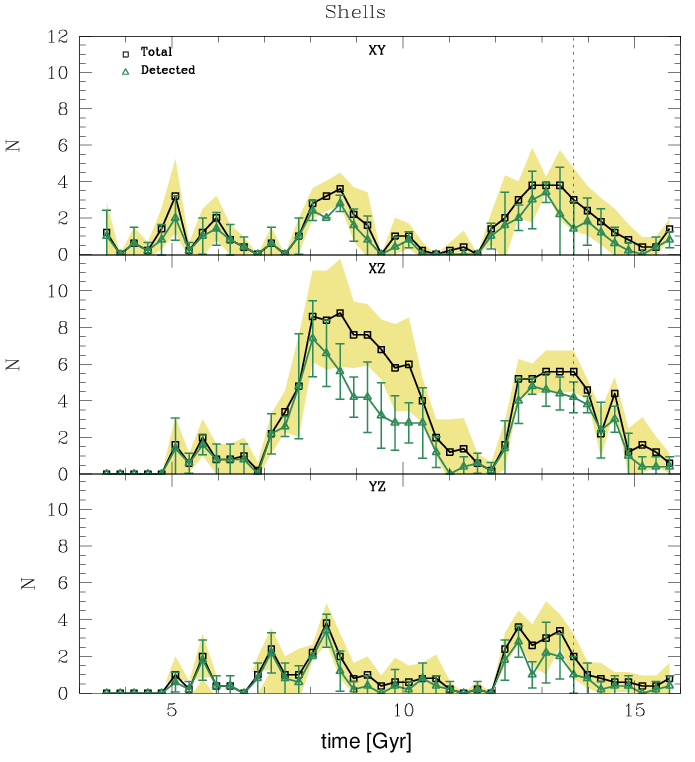}
\end{tabular}
\caption{Evolution of shell numbers.  Three panels show
     results for three projections. In each panel, the black lines
        and the yellow shaded area are
    the total number of shells and their respective scatter, 
        whereas the green ones represent the
    number of shells that are detectable on images with a cut in
    surface brightness at 29 mag arcsec$^{-2}$.}
\label{evoshells}
\end{figure}

This class of substructures is highly dependent on the projection effects. 
The middle panel of figure \ref{evoshells}
(XZ-projection) demonstrates this result in particular. The incidence of shells is much 
higher in this projection than in the other two orientations. In Figure \ref{shell} can be appreciated 
shells in only one orientation.
The explanation is that shells are aligned with each other, either along one axis (prolate case)
or a common equatorial plane (oblate case) of the central galaxy \citep{Dupraz1986}.



\section{Discussion}
Table \ref{tab2} summarizes some of our findings and the characteristics
of the three classes of fine structures.
Colors are not considered in this paper given the difficulty to derive them in real observations, although they are provided 
in the simulations.
With the deeper surface brightness limit, it is possible to detect shells in the most
favorable orientation up until 4 Gyr, stellar streams when they are not destroyed by
further minor mergers and until 3 Gyr;  and tidal tails up until 2 Gyr. Tidal tails are considered the main tracer of major merger but they vanish more quickly.
Streams are thinner and more frequent since they are produced in minor mergers, which have
a much greater occurrence. Shells are highly dependent on sky orientation and they are the oldest vestiges of intermediate-mass mergers. In addition, their number increases
over time, although their surface brightness decreases simultaneously.

It is interesting to compare these results with previous studies. 
\citet{Lotz2008} and \citet{Ji2014} identify different timescales for different 
merger stages of their simulated pair encounters.
From their series of hydrodynamical simulations of equal-mass
gas-rich disc galaxy mergers, \citet{Lotz2008} find characteristic perturbed
and asymmetric morphologies during the merging time, but quite relaxed ones
afterwards. Their observability timescales for the resulting fine structures
are dependent on  numerous parameters (viewing angle, orbit geometry, gas fraction, etc.)
but are typically found in the same order as the coalescence timescale. 
The remains of the intense
starburst
triggered by the major merger is more easy to see than morphological perturbations. Their merger remnants appear disk-like and dusty, while exhibiting a massive
bulge. Our simulation is of a different character as it follows a series
of minor mergers and gas accretion in a cosmological context that is undergone by 
a central galaxy which is become an early-type quenched object. The observability
timescales are therefore more realistic, taking into account successive
mergers which are capable of destroying the vestiges of previous ones.

 \citet{Ji2014} also follow a series of mergers, but their study is based on different
 mass ratios and use hydrodynamical simulations. They consider the survival timescales of the 
 fine structures according to the surface brightness limit and galaxy environment.
When isolated, the survival time is, on average, twice the coalescence time for a
surface brightness limit of 25 mag arcsec$^{-2}$, i.e. $\sim$ 2.5 Gyr. This timescale
can be twice as long ($\sim$ 5 Gyr) for a deeper surface brightness limit of 28 mag arcsec$^{-2}$.
 Again, our simulation is more realistic since
the central galaxy is the result of several satellite mergers
during its formation and evolution, with different mass ratios ranging from
25:1 to 1.5:1, making the system more chaotic.
For a minor merger of a small mass ratio, the final fine structure timescale can be quite long
since the coalescence timescale itself is longer than for a major merger.
The occurrence of minor mergers is traced more frequently by stellar streams than by
tidal tails (which tend to characterize major mergers). 

The advantages of our method, relative to the systematic study of individual mergers, is
that we can study the survival time of fine structures based on a complex merging history.
The results are then directly comparable to observations because, in addition, we use the
same classification method sensitive to LSB features.
  To counterbalance these advantages, there are obvious limits to our method: we
focus on only one merger history, and more computational efforts will be required
  to obtain a wider set of statistics. 
Also, it is difficult to follow individual substructures from one snapshot to the other: we only count  the total number of features,
and there is some degeneracy when one feature is destroyed and another one created, resulting in
the same number. To reduce the degeneracy, more frequent snapshots should be documented,
which would prove more demanding for the process of visual classification.

\begin{table}
   \caption{Summary of results}
\begin{tabular}{ |l|l|l|l| }
\hline
\multicolumn{4}{ |c| }{Properties of fine structures} \\
\hline
Property         & Tidal tails  & Shells        & Streams\\ \hline
  & -1st \& 3rd peak:   & -1st peak:    & -1st peak:   \\
Formation   &  Mass ratio (1:8)&   ratio (1:8)& ratio (1:3) \\
   &            &               & \\
 & -2nd \& 4th peak:    & -2nd peak:    & -2nd peak: \\
mechanism   &    (1:3, 1:1.55)&  ratio (1:1.55)& ratio (1:8) \\ 
   &                    &       & \\ \hline
Visibility  &  0.7-1 Gyr& 3-4 Gyr       & 1.5-3 Gyr\\
\hline
 Surface    &  & &   \\
 brightness   & No & Mild&  High \\
 sensitivity   &  & &   \\
        \hline
 Orientation      & No  & High  & No   \\ 
 dependency       &   &   &   \\ 
        \hline
\multirow{2}{*}{Color (B-V)}   & Blue: & Red:           & Blue:  \\
  & -2.5 to -1.5 &  -1.5 to 0   &  -3 to -1  \\
\hline
\end{tabular}
\label{tab2}
\end{table}

We demonstrate how fine structures in the outer parts and stellar halos of galaxies
can help to trace back the merger history and the mass assembly of present day galaxies.
A study of stellar halos of Milky Way-like galaxies in the nearby universe, compared with
Illustris simulations, have shown that the morphology of galaxies are, indeed, closely linked
to the significance of their halos \citep{Elias2018}.
Galaxies which demonstrate a small stellar halo mass relative to their total mass are disk galaxies which
are still forming stars, whereas those with a large fraction of their mass in their halos are quenched.
Simulations by \citet{Karademir2019} show that minor mergers enrich the stellar halos
of galaxies much more than their centers and that they are able to increase the size of galaxy disks,
as has been observed \citep[e.g.,][]{Newman2012}.

In recent years, machine learning algorithms have been developed to identify
fine structures in galaxy halos, both in observations and simulations
\citep[e.g.][]{Walmsley2019, Hendel2019}. Comparisons to visual classifications
provide promising results and these methods will be inevitable for  large surveys in the future.

\citet{Mantha2019}  propose a tool for identifying tidal debris in CANDELS deep
fields and  in the VELA simulations as well, which helps to disentangle viewing effects
and  the subjectivity of visual inspections. Automatic tools like this will help
to improve the characterization of the merger rate as a function of redshift.

\section{Conclusions}
 We used a hydrodynamical simulation that has been resimulated within a cosmological 
 context in order  to analyze and interpret the morphologies and survival timescales
 of fine structures, as well as tracers of mergers and of the mass assembly of present day
 early-type galaxies. We took a census of three types of fine structures based on visual inspection:
 tidal tails, stellar streams, and shells. The observation of the number of fine structures detected 
 around the central galaxy versus time allows us to reconstruct its merger history. 
We find three phases in the central galaxy evolution, the second one characterized as a very quiet
phase.  The census of fine structures are taken in correspondence with the 
 merger events that have been identified. This allows us to establish that:

\begin{itemize}
\item
Tidal tails result from major mergers events (1.5:1), stellar streams from minor mergers (10:1) 
and shells from major and intermediate-mass mergers (4:1).
\item
Tidal tails and shells have long survival times, $\sim$2 Gyr and $\sim$3Gyr, respectively, but 
streams remains visible across all phases of galaxy evolution.
\item
The detection of stellar streams are highly dependent on the surface brightness limit.
We see between two and three times more streams with a surface brightness cut of 33 mag arcsec$^{-2}$ than with
29 mag arcsec$^{-2}$.
\item
  The detection of shells  depends  considerably on the projection angle and orientation on the sky plane.
\end{itemize}

  Our results are compatible with previous simulations of comparable scope, however, conducting the zoom-in
  resimulation in a cosmological context introduces more realistic conditions to estimate
  the survival timescales of fine structures around today's early-type galaxies.

  As for the observer's point of view, the reconstruction of the merger history is 
  obtained more optimally with shells, which have longer timescales. The number of shells can also
  indicate the time that has passed since the last merger. Streams have shorter timescales, so when no fine
  structure is observed, this means either no event has occurred since 2 Gyr, or it has been limited to a very minor merger with
  a large angular momentum, which produces more streams than shells.

\begin{acknowledgements}
  We thank the anonymous referee for their constructive comments,
and David Valls-Gabaud for the helpful discussion.
B. Mancillas was supported in part by the CONACYT grant, CVU 420397.
\end{acknowledgements}




 




%
\appendix
\section{Snapshot Evolution}
\begin{figure*}
\centering
\begin{tabular}{c}
\includegraphics[ scale=3.0, trim=30 0 0 0,clip]{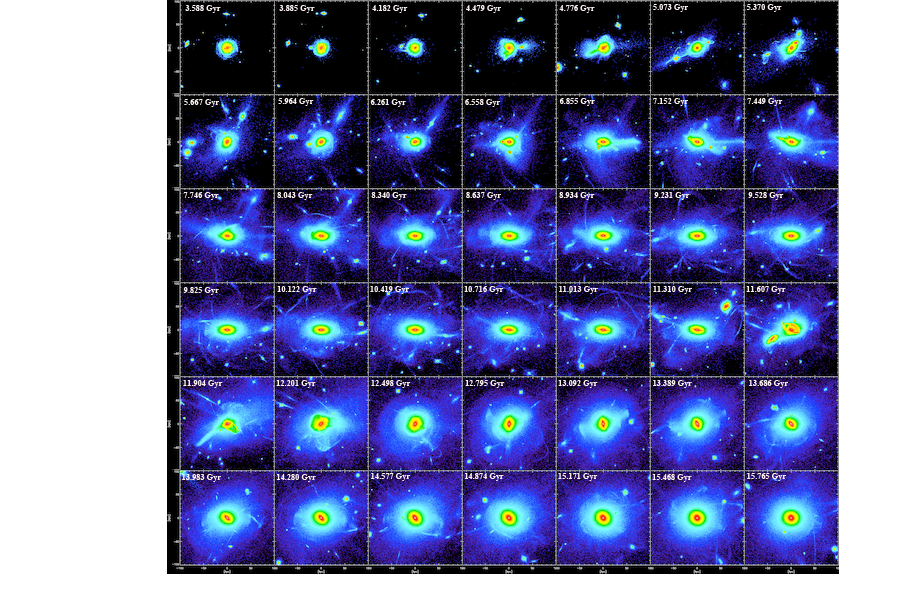}
\end{tabular}
\caption{Snapshot evolution in XY-Projection.}
\label{snapXY}
\end{figure*}

\begin{figure*}
\centering
\begin{tabular}{c}
\includegraphics[ scale=3.0, trim=30 0 0 0,clip]{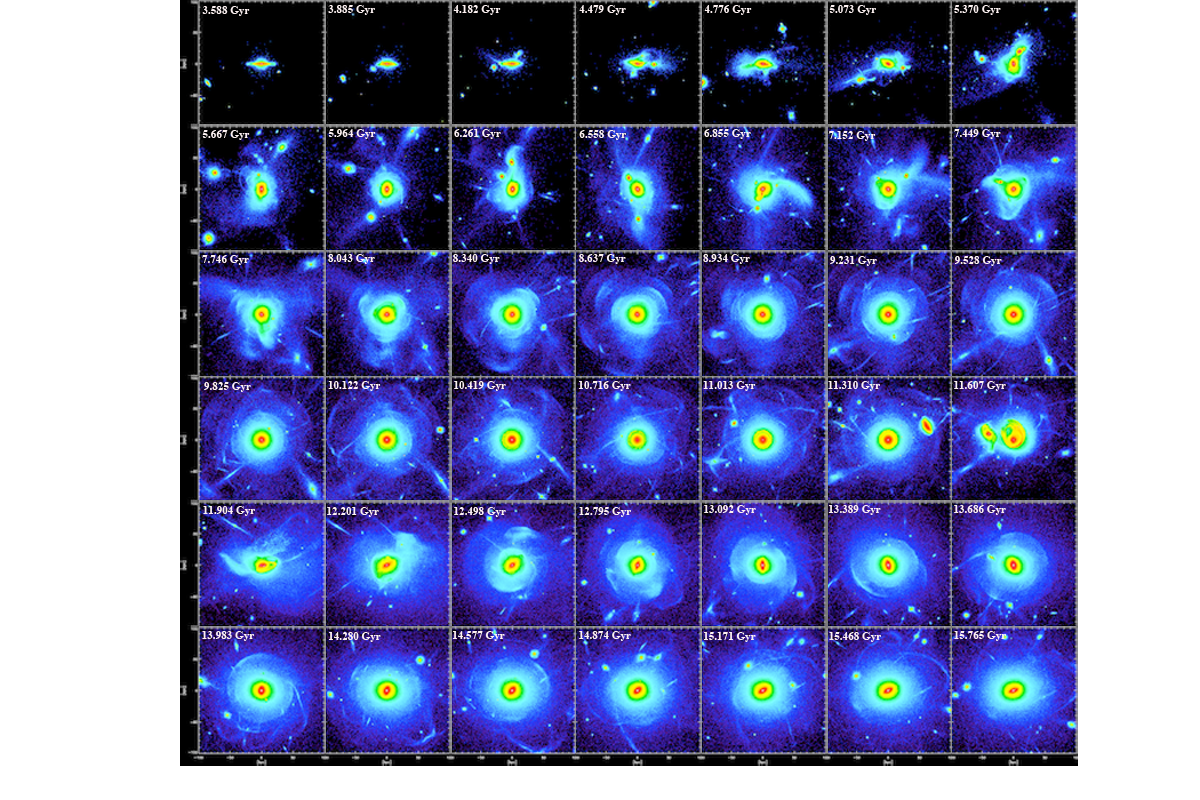}
\end{tabular}
\caption{Snapshot evolution in YZ-Projection.}
\label{snapXZ}
\end{figure*}

\begin{figure*}
\centering
\begin{tabular}{c}
\includegraphics[ scale=3.0, trim=30 0 0 0,clip]{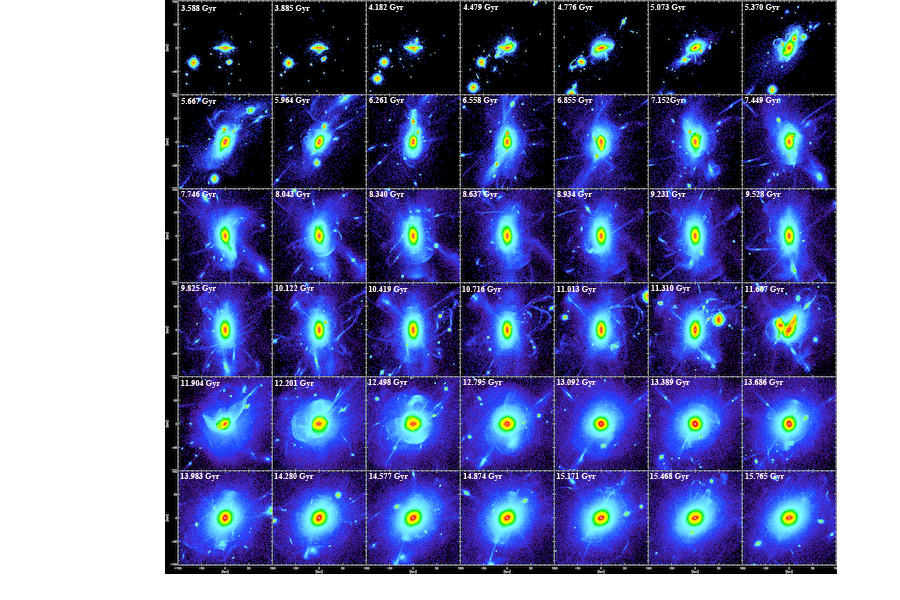}
\end{tabular}
\caption{Snapshot evolution in the XZ-Projection.}
\label{snapYZ}
\end{figure*}

In the following three figures (Fig. \ref{snapXY}, \ref{snapXZ} and \ref{snapYZ}),
we present the whole ensemble of snapshots of the zoom-in simulation of the 3 projections.
%
%
%
%
\newpage

\end{document}